\documentclass[aps,pra,floatfix,amsmath,ams,preprint,superscriptaddress,showpacs]{revtex4}
\usepackage{graphicx}
\usepackage{dsfont}
\begin{document}

\title{The modular multiplication operator and the quantized bakers maps}

\author{ Arul Lakshminarayan\footnote{On leave from: Department of Physics, Indian Institute of Technology Madras, Chennai, 600036, India.}}
\email[]{arul@physics.iitm.ac.in}
\affiliation{Max-Planck-Institut f\"ur Physik komplexer Systeme, N\"othnitzer Stra$\beta$e 38, D-01187 Dresden, Germany}
\date{\today}
\preprint{IITM/PH/TH/2007/7}

%\date{\today}
\begin{abstract}
The modular multiplication operator, a central subroutine in Shor's factoring algorithm, is shown to 
be a coherent superposition of two quantum bakers maps when the multiplier is $2$.  The
 classical limit of the maps being completely chaotic, it is shown that there exist perturbations that push the modular multiplication operator into regimes of generic quantum chaos with spectral fluctuations that are those of random matrices. For the initial state of relevance to Shor's algorithm we study fidelity decay due to phase and bit-flip errors in a single qubit and show exponential decay with shoulders at multiples or half-multiples of the order. A simple model is used to gain some understanding of this behavior.
%\pacs{05.45.Mt,03.67.Lx}
\end{abstract}
\maketitle

%\newpage

\newcommand{\newc}{\newcommand}
\newc{\beq}{\begin{equation}}
\newc{\eeq}{\end{equation}}
\newc{\kt}{\rangle}
\newc{\br}{\langle}
\newc{\beqa}{\begin{eqnarray}}
\newc{\eeqa}{\end{eqnarray}}
\newc{\pr}{\prime}
\newc{\longra}{\longrightarrow}
\newc{\ot}{\otimes}
\newc{\rarrow}{\rightarrow}
\newc{\h}{\hat}
\newc{\bom}{\boldmath}
\newc{\btd}{\bigtriangledown}
\newc{\al}{\alpha}
\newc{\be}{\beta}
\newc{\ld}{\lambda}
\newc{\sg}{\sigma}
\newc{\p}{\psi}
\newc{\eps}{\epsilon}
\newc{\om}{\omega}
\newc{\mb}{\mbox}
\newc{\tm}{\times}
\newc{\hu}{\hat{u}}
\newc{\hv}{\hat{v}}
\newc{\hk}{\hat{K}}
\newc{\ra}{\rightarrow}
\newc{\non}{\nonumber}
\newc{\ul}{\underline}
\newc{\hs}{\hspace}
\newc{\longla}{\longleftarrow}
\newc{\ts}{\textstyle}
\newc{\f}{\frac}
\newc{\df}{\dfrac}
\newc{\ovl}{\overline}
\newc{\bc}{\begin{center}}
\newc{\ec}{\end{center}}
\newc{\dg}{\dagger}
\newc{\md}{\mbox{mod}}
\newc{\prq}{\mbox{PR}_q}
\newc{\rt}{\sqrt{2}}

\section{Introduction}

Given an $M$ dimensional complex Hilbert space consider an orthonormal basis $|m\kt$, $m=0,\ldots,M-1$. 
The modular multiplication by a number $p$ coprime to $M$ is the unitary permutation operator $U_p$:
\beq
U_p|m\kt \longrightarrow  |m\, p \,( \mbox{mod}\, M)\kt.
\eeq
Repeated application of $U_p$  is also the modular exponentiation operator and lead to $U_p^k|m \kt =|m \,p^k\, ( \mbox{mod}\, M)\kt$. $U_p$ is periodic, i.e. there exists $k_0$ such that $U_p^{k_0}=I$, or $p^{k_0}= 1(\md M)$. This period though is an irregular function of $M$ for a given $p$ and is the ``multiplicative order''  of $p\, (\mbox{mod}\, M)$ \cite{Mathworld}. This operator is a crucial subroutine of Shor's factoring algorithm \cite{Shor} in which it is required to 
perform the following operation on a bipartite Hilbert space: 
\beq
|j\kt |1\kt \longrightarrow |j \kt | p^j \md M \kt
\eeq
where $0\le j \le 2^t-1$ and $M<2^t$. Typically $M$ is a large number and therefore we need
to calculate large powers and their residues $\md M$. This is considerably simplified by modular
arithmetic \cite{NielsenChuang} and the whole modular exponentiation step can be performed with $\mathcal{O}((\log M)^3)$
number of gates \cite{Beckman1996}. Once this is done, a quantum Fourier transform extracts the 
period $k_0$ with reasonable rate of success. Given the period (order) it is possible to find a factor efficiently, 
provided that $p_{k_0}$ is even and $p^{k_0/2} \neq -1 (\md M)$ by well-known procedures of number theory, 
and using only classical computers \cite{NielsenChuang}.

Shor's algorithm exploits the polynomial speed of the quantum  Fourier transform to find the order
and hence offers a polynomially scaling algorithm for factoring numbers. Shor's algorithm has been implemented experimentally \cite{Chuang} although the number so factored is still very small to excite any practical application. The
effect of decoherence and gate errors on Shor's algorithm are important considerations and have been addressed by several authors previously \cite{Chuang96,Fowler,Hu,Devitt04,Miquel,Shepel07}. For instance Ref.~\cite{Chuang96}discuss the impact of environmental decoherence on the algorithm, while in \cite{Devitt04} direct detailed simulations have shown that the Shor algorithm is highly sensitive to gate errors, and the effect of  static imperfections have been studied recently in\cite{Shepel07}. In this paper we will not directly simulate Shor's algorithm but look closely at the 
modular multiplication and exponentiation for the special and simplest case $p=2$. We will call
$U_2$ as $S$, the shift operator as it performs the simple action of a qubit cyclic shift if $M$ is a power of $2$.

Quantum algorithms have  been studied earlier with a view to see if they had properties of 
quantum chaotic systems \cite{Braun,Maity}. Recently it was shown that the spectrum of the unitary part of Shor's algorithm, properly desymmetrized had typical random matrix fluctuations \cite{Maity}, indicating that the operator itself may be quantum chaotic. However it was also pointed out that the origin of the chaos is the modular exponentiation part which is akin
to nongeneric quantum chaotic systems such as the cat maps \cite{QCat}. Here we make this connection more
precise and show that the classical limit of these subroutines is an ad-mixture of  two bakers maps. Bakers maps are paradigms of deterministic classical chaos that are as random as a coin toss \cite{LL,Tabor}. The dimensionless inverse Planck constant  in the Shor algorithm is the number to be factored and hence the classical limit is reached through a practically important regime. 
We show that due to the proximity with such operators, there are perturbations that push the modular exponentiation
part (and therefore indeed the whole of Shor's algorithm) into regimes of generic quantum chaos. However we are not
that much interested in stationary state properties as in time-evolving states, in fact on those states that are 
used in Shor's algorithm. Hence we study the fidelity of repeated modular multiplication, or the modular exponentiation, and show how the decay depends on the classical limit. We also provide a simple model for the exponential fidelity decay that is exactly solvable and captures the actual behavior reasonably well.

\section{The baker and the shift operator}

The classical baker's map\cite{LL,ArnAvez,Tabor} $B_c$ is an area-preserving transformation
of the unit phase-space square $[0,1)\times [0,1)$ onto itself, which takes a
phase-space point $(q,p)$ to $(q',p')$ where $(q'=2q,\, p'=p/2)$ if
$0\le q<1/2$ and $(q'=2q-1,\,p'=(p+1)/2)$ if $1/2\le q<1$.  The
stretching along the horizontal $q$ direction by a factor of two is
compensated exactly by a compression in the vertical $p$
direction. This is well known to be a fully chaotic system that in a
mathematically precise sense is as random as a coin toss\cite{Ornstein}. The
area-preserving property makes this map a model of chaotic two-degree
of freedom Hamiltonian systems. The lack of a generating Hamiltonian is compensated
by the existence of a classical generating function of the canonical transformation $B_c$.
 The chaos is inferred by expressing a phase space point in the binary representation, if
$q=0.a_0a_1a_2\cdots$ and $p=0.a_{-1}a_{-2}a_{-3}\cdots$, where $a_i$
are either $0$ or $1$, then $q'=0.a_1a_2\cdots$ and
$p'=0.a_{0}a_{-1}a_{-2}a_{-3}\cdots$.  Thus the most significant bits
of $q$ are lost at the rate of a bit per iteration, leading to an
exponential increase in any initial error. The Lyapunov exponent is
$\log(2)$ per iteration.  This ``left-shift'' is in fact an important
mechanism for the generation of Hamiltonian chaos\cite{Tabor}, and in more
complicated forms arises generically.

This was quantized first by Balazs and Voros\cite{BalVor1,BalVor2}. 
 Quantization in this context is the construction
of an appropriate unitary operator that evolves states over one
iteration and has the correct  classical limit. Symmetries that may be broken
on quantization must be restored in this limit.  The
Hilbert space of states is finite dimensional, has $N$ position and
momentum states, denoted by $|q_n\kt$ and $|p_m\kt$. If periodic
boundary conditions are assumed, $|q_{n+N}\kt=|q_n\kt$,
$|p_{m+N}\kt=|p_m\kt$, this implies that the transformation functions
between position and momentum is the discrete Fourier transform:
$(F_{N})_{mn}=\br p_m|q_n\kt = \exp[-2 \pi i mn/N]/\sqrt{N}$,
$m,n=0,1,2,\ldots,N-1$. Here $N$ is an effective scaled Planck
constant as $N=A/h=1/h$, where $A$ is the area of the phase space,
here unity. Thus the classical limit is the large $N$ limit. If $B$ is
the quantum baker's map, Balazs and Voros required that $\br
p_m|B|q_n\kt = \sqrt{2}\, \br p_m|q_{2n}\kt\, =\, (F_{N/2})_{mn}$ if
$n$ and $m$ are {\it both} $\le N/2-1$. This is almost like requiring
that $B$ takes $|q_n\kt$ to $|q_{2n}\kt$ mimicking the classical
stretching action, except that the momentum components above $N/2$ 
are set zero. ($\br p_m|B|q_n \kt=0$ for $p_m\ge N/2$ and $q_n <N/2$).
 It is also clear from this that $B$ is very close to the action of 
modular multiplication with $p=2$ \cite{aruljphys}. $N$ is throughout assumed to
be an even integer. In fact we will set $N=2L$ and can then consider the
quantum baker to act on a Hilbert space of a qubit coupled to an $L$ dimensional
systems. 

A similar argument is made for the second half of the transformation, and remarkably these conditions are 
consistent and produce an unitary operator which has a broken parity symmetryc \cite{BalVor2}.
The classical symmetry being $(q \longrightarrow1-q, \, p\longrightarrow 1-p)$.  Saraceno\cite{Saraceno} 
restored this by imposing anti-periodic boundary conditions, and this leads
to the quantum baker's map:
\beq B=G_{2L}^{-1}\left( \begin{array}{cc}G_{L} &0
\\0 & G_{L} \end{array}\right), \eeq where $(G_{N})_{mn}=\br
p_m|q_n\kt = \exp[-2 \pi i (m+1/2)(n+1/2)/N]/\sqrt{N}$. 
$B$ is an unitary matrix, whose repeated application
is the quantum version of the full left-shift of classical chaos. 
This quantum map has been continued to be studied as it has many properties
of generic quantum chaotic systems, including random matrix like
spectral fluctuations \cite{BalVor2} and eigenfunction scarring \cite{Saraceno}. It is also amenable to 
a simple semiclassical periodic orbit sum, and hence has been used in the 
study of such approximations \cite{SarAlm,Connor}. For $N$ that are powers of $2$ it
was found that the Hadamard and related transforms highly simplified the
eigenstates and some of them are remarkably well described by the Thue-Morse
 sequence \cite{Schroeder} and its Fourier transform \cite{MeenArul12,Ermann}. It has also been used in the study of entanglement \cite{ScottEnt} and hypersensitivity of quantum chaos \cite{Scott07}. It is possible to design a quantum
circuit for the quantum baker map \cite{SchackQBM} and this been implemented on a NMR  quantum computer
experimentally \cite{QBMexp}.

If one is not mindful of classical
symmetries being fully preserved on quantization, there are a large number of possible
quantum baker maps \cite{BalVor2,SarVor}. An important class of such ``decorated bakers'' \cite{Voros} are got 
by embellishing the original bakers map with relative phases in the half-sized Fourier
blocks, as well as in the definition of the Fourier transform itself, as done below.
In a previous work we have constructed such a decorated quantum
bakers map using the shift operator $S$ \cite{aruljphys}. It will be useful to do
the converse and construct the shift operator from the quantum 
bakers map or similar operators. It is well-known that the ``hard'' part of
Shor's algorithm is the implementation of the modular exponentiation
step. On the other hand the quantum bakers map is implemented with 
quantum Fourier transforms (QFTs) and this may make the implementation of
the shift operator possible with the QFTs. We explicitly show this at least for
the case $p=2$. More importantly for us it will enable embedding the shift operator in a larger family of operators which includes maps with well defined classical limits, thereby making the classical limit
of modular multiplication explicit. 

The shift operator we have already defined, however we restate it for clarity as:
\beq
S|n \kt =|2 n (\md N-1) \kt
\eeq
with the caveat that $S|N-1\kt =|N-1\kt$. This corresponds to our earlier definition with 
$M=N-1$ with one more state ($|N-1\kt$) added to the Hilbert space, but which remains fixed, and 
outside any dynamics we are interested in, but may participate when there are perturbations.
Note that since for the bakers map $N$ is an even integer $S$ is unitary.
We define a generalized Fourier transform as
\beq
(F_{N}(\al,\beta))_{nm}=\df{1}{\sqrt{N}} \exp(-2 \pi (n+\al)(m+\beta)/N)
\eeq
Evaluating the product of the  Fourier transform and $S$ we can derive, merely by summing finite geometric
series and using elementary properties of exponentials that
\beq
S=\dfrac{1}{\sqrt{2}} F_{2L}^{-1}(\al,\al) \, \left( \begin{array}{cc} F_{L}(\al,\f{\al}{2})&F_{L}(\al,\f{1+\al}{2})\\
e^{-i \pi \al} F_{L}(\al,\f{\al}{2})& -e^{-i \pi \al} F_{L}(\al,\f{\al+1}{2}) \end{array} \right).
\eeq
Note that the operator $S$ does not depend on the phase $\al$ that appears on the R.H.S.. We use
this freedom to break or keep the parity symmetry. A natural and simple choice is $\al=0$, but
$\al=1/2$ leads to symmetric operators as explained below. The structure of the above identity
allows this to be written as
\beq
S=\df{1}{\sqrt{2}}\left( B_{2L} \, +\, B'_{2L}\right)
\eeq
where 
\beq
B_{2L}=  F_{2L}^{-1}(\al,\al) \, \left( \begin{array}{cc} F_{L}(\al,\f{\al}{2})&0\\
0& -e^{-i \pi \al} F_{L}(\al,\f{\al+1}{2}) \end{array} \right).
\eeq
and
\beq
B'_{2L}= F_{2L}^{-1}(\al,\al) \, \left( \begin{array}{cc} 0 &F_{L}(\al,\f{1+\al}{2})\\
e^{-i \pi \al} F_{L}(\al,\f{\al}{2})& 0 \end{array} \right).
\eeq

\begin{figure}
\includegraphics[height=4in]{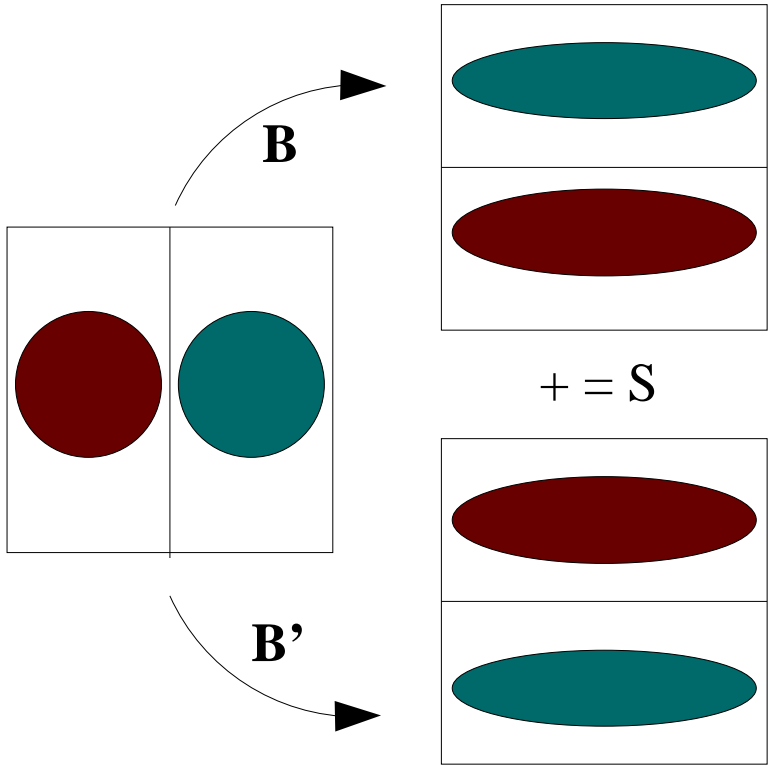}
\caption{A schematic view of the shift operator as a sum of two bakers maps, one 
with the usual stacking order of the vertical left half being stretched to the bottom horizontal half ($B_c$-classical, $B_{2L}$-quantum)
and the baker with a reverse stacking order ($B^{'}_{c}$-classical, $B'_{2L}$-quantum). }
\label{bakers}
\end{figure}  

Thus remarkably the modular multiplication $S$ can be written as a {\it sum} of two
 unitary operators (with a normalization factor). $B_{2L}$ and $B'_{2L}$ are two {\em quantum baker maps}, of which the former one is the standard one, which we have discussed above. It is well known that such decorated bakers also 
perform the same classical actions as normal bakers \cite{BalVor2,SarVor}, the classical limit being 
$L \longrightarrow \infty$. The operator $B'_{2L}$ has not been studied nearly as much,
but has recently appeared in a work that uses this to study coupled chaotic systems \cite{Vallejos07}.
The classical limit (say $B'_c$) as pointed out in this study corresponds to a different stacking
order of the vertical partitions of the bakers map after they have been stretched. 
Instead of the usual left-half transiting to the bottom-half, it is put in the top-half
and the right-half goes into the bottom half. This fixes the lower right-hand corner
of the square. Thus $B'_c(q,p)=(2q,(p+1)/2)$ if $q\le 1/2$ and $(2q-1,p/2)$ if $q>1/2$.
Again the operator $B'_{2L}$ that appears above differs from the 
one used earlier in terms of the ``decorations''. Of course these decorations are 
absolutely crucial so that the two unitary evolutions, which are non-periodic and 
have random matrix like properties,  {\em add} and conspire to produce the simple shift operator that is completely periodic. Previous studies of the classical limit of operators such as $S$ include those
of what is called the ``extremal quantum baker map'' \cite{SchackCaves} and it has been suggested that the classical limit  corresponds to a  ``stochastic classical map'' \cite{Scott,Nonnen}. In Fig.~(\ref{bakers}) we have shown a schematic of the classical bakers maps that on quantization and coherent addition yield the shift operator. Also see Ref.~\cite{aruljphys} for a 
description and figure of the action of $S$ on Weyl coherent states.

That the simple shift operator's can be thought of as a coherent superposition of 
two quantum chaotic evolutions has been demonstrated above in a particularly simple way.
This suggests that there maybe perturbations of the operator $S$ that are generic and may possess
random matrix \cite{Mehta} like and other quantum chaotic properties \cite{GutzBook,HaakeBook}.
 We show below that this is indeed the case. Since quantum chaotic operators also are typically sensitive to perturbations
\cite{CavesSchack,Peres}, this may have implications for the operation of the Shor algorithm. We partly study this by measuring the fidelity of $S$ to small perturbations and show that the fidelity decays exponentially in time till the order of $N-1$ or half of this. Thereafter it typically shows enhanced rate of decays at multiples of this time, but could also show strong recurrences. Surprisingly a simple analysis when $N$ is a power of 2 captures many of the qualitative features of the more general case.

\section{Perturbations of the shift operator and quantum chaos}

In terms of operations on the Hilbert spaces of the tensor product ${\cal H}_2 \otimes {\cal H}_L$
we may write the shift operator as
\beq
S= F_{2L}^{-1}(\al,\al)\circ\df{1}{\sqrt{2}} \left( \begin{array}{cc}  1 & 1 \\ e^{-i \pi \al} & -e^{-i \pi \al} \end{array}\right) \otimes I_L \circ
\left( \begin{array}{cc}  F_L(\al,\f{\al}{2}) &0 \\0 & F_L(\al,\f{1+\al}{2}) \end{array} \right).
\eeq
Thus the modular exponentiation maybe implemented with QFTs that have suitable phases. However the dimensionalities
of the QFTs are not in general powers of 2 and are therefore not the standard ones in use.
We choose to perturb the central operator in the above equation, and perturb only the 
qubit space ${\cal H}_2$. In particular we consider the smooth embedding of the shift
operator in the family:
\beqa
S(\theta; \al,P)&=& F_{2L}^{-1}(\al,\al)\circ \exp(-i \theta P) \df{1}{\sqrt{2}} \left( \begin{array}{cc}  1 & 1 \\ e^{-i \pi \al} & -e^{-i \pi \al} \end{array}\right) \otimes I_L \circ
\left( \begin{array}{cc}  F_L(\al,\f{\al}{2}) &0 \\0 & F_L(\al,\f{1+\al}{2}) \end{array} \right) \nonumber \\
&=&
F_{2L}^{-1}(\al,\al) \circ \exp(-i \theta P) \otimes I_L \circ  F_{2L}(\al,\al) \, S \, = \,V(\theta) S.
\eeqa
here $P$ is the perturbing Hermitian operator on the qubit space, and $V(\theta)$ defined through the
last equation is the Fourier transform of the perturbation generated by $P$.  The operator $S(0;\f{1}{2},P)$
is the unperturbed shift operator simply called $S$ so far. The family of operators $S(\theta;\al,P)$ now depends on the phase $\al$ as well, although $S(0;\al,P)$ does not. We display this dependence explicitly as the 
phase $\al$ does play a crucial role.

The operator $S$ has the quantum parity symmetry $R$ : $R|n \kt = |N-n-1\kt$, that is 
$SR=RS$. If $L$ is a power of 2, then $R$ is simply the product $\otimes ^{2L} \sigma_x$.
Perturbations of the shift will therefore in general approximately preserve this symmetry. To analyse
 random matrix properties it is desirable to completely break a symmetry or preserve it and
desymmetrize the operators. Since we want to retain the character of a small perturbation, we
first choose to preserve the parity symmetry exactly. We can do this by adopting anti-periodic
boundary conditions, $\al=1/2$ and choosing $P=\sigma_x$. This will lead to the family:
\beq
\label{alhalfX}
S(\theta; \f{1}{2},\sigma_x)=F_{2L}^{-1}(\f{1}{2},\f{1}{2}) \left( \begin{array}{cc} \sin(\f{\pi}{4}-\theta) F_{L}(\f{1}{2},\f{1}{4})& 
\cos(\f{\pi}{4}-\theta) F_{L}(\f{1}{2},\f{3}{4}) \\-i\cos(\f{\pi}{4}-\theta) F_{L}(\f{1}{2},\f{1}{4})&
i\sin(\f{\pi}{4}-\theta) F_{L}(\f{1}{2},\f{3}{4})
\end{array} \right).
\eeq
When $\theta=\pm \pi/4$, the operators correspond to the bakers of type $B'_{2L}$ and $B_{2L}$
respectively.  For other angles it represents a coherent mixture of the two types of baker operator stacking
while at  $\theta=0$ it is the usual shift operator.

\begin{figure}
\includegraphics[height=5in,angle=-90]{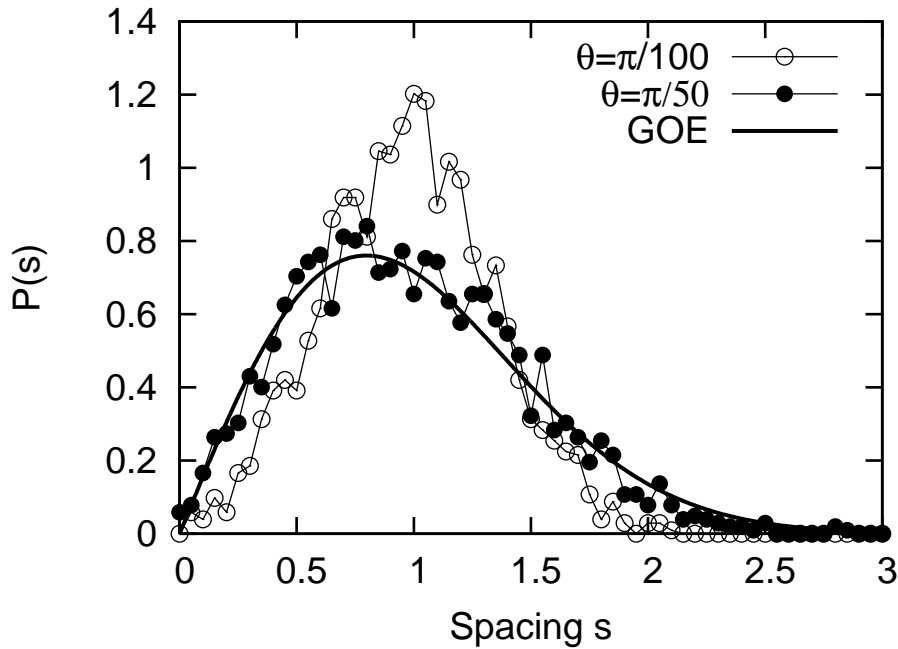}
\caption{The nearest neighbor spacing distribution of even-subspace eigenangles 
for two perturbations of the shift operator when  $N=4094$, the perturbations
preserving the parity symmetry.  The smooth curve  shows the corresponding 
GOE result of random matrix theory, the Wigner-Dyson distribution.
}
\label{nns}
\end{figure}

We take the even subspace of the spectrum of $S(\theta;\f{1}{2},\sigma_x)$ and show in 
Fig. ~(\ref{nns}) the nearest neighbor spacing statistics for two case of small angles $\theta$.
It is clear that if the perturbation is very small, the rigid, harmonic oscillator like spectrum
of $S$ widens into one where there is dominant level repulsion, and for fairly significant 
perturbations becomes a generic one that belongs to the universality class of the
Gaussian Orthogonal Ensemble (GOE) of random matrices \cite{Mehta}, well-known to apply
to quantum chaotic systems that have time-reversal symmetry \cite{Bohigas84,HaakeBook}. 
Previously it was shown that the unitary part of the full Shor algorithm, including the Hadamard and the
Fourier transforms, had fluctuations that were of the CUE kind \cite{Maity}. We note here that restricting ourselves
to the modular multiplication part with a particular perturbation allows us to preserve the
time-reversal that holds for individual quantum bakers maps \cite{BalVor2}.

Thus we see that indeed there are perturbations of the shift-operator that are quantum chaotic. There is also a crucial 
dependence on the number $N$ (or the number to be factored $N-1$). If $N$ were a power of
$2$ such as $4096$, instead of $4094$ in Fig. ~(\ref{nns}) there would be much more 
deviation from the GOE distribution, with a large peak near the origin. This anomalous
statistic arises from the extreme degeneracy of the eigenangles when $N$ is a power of 2,
and is special. A similar situation where there is an extreme dependence of the statistics
of the spectrum on the effective Planck constant $N$ arises in the case of the perturbed
cat maps \cite{Esposti05}, and presumably for similar reasons. Earlier it has also been pointed out that perturbing the
cat maps slightly so that the sawtooth map arises leads to a rapid restoration of the generic
fluctuation characteristics of quantum chaotic systems \cite{arulbalazssaw}. Thus the similarities of the shift map
to the quantum chaotic cat maps with their special arithmetic properties is further highlighted here.

\section{Fidelity decay}

We turn to non-stationary properties, as indeed the Shor algorithm is the result of time evolution of
a particular initial state which corresponds to the state $|1\kt$. The algorithm requires requires finding
 the states $|x^j \, \mod(N-1)\kt$ for $x$ coprime to $N-1$. As stated previously  we take  $x=2$ throughtout,
and we now study how gate errors would proliferate in time. In particular we study the fidelity
\beq
f(t)= |\br 1 |S^{-t} S^t(\theta; \al,P) |1\kt |^2\, =\, |\br 2^t\,  \mbox{mod}(N-1)|S^t(\theta; \al,P)|1\kt |^2
\eeq
In this section choose $\alpha=0$ for simplicity and note that there is a weakly broken
parity symmetry as a result of this. 

\subsection{Case: $P=\sigma_x$}

The first case we take will be a rather special one wherein
the perturbation is ineffective: the bit-flip $P=\sigma_x$.  In this case we have that 
\beq
S(\theta; 0,\sigma_x)=F_{2L}^{-1}(0,0) \left( \begin{array}{cc} \f{1}{\rt}e^{-i \theta} F_{L}(0,0)& 
\f{1}{\rt}e^{i \theta} F_{L}(0,\f{1}{2}) \\ \f{1}{\rt}e^{-i \theta} F_{L}(0,0)&
 \f{1}{\rt}e^{i \theta}  F_{L}(0,\f{1}{2})
\end{array} \right).
\eeq
and $f(t)=1$ for all time $t$. Note that there are only phases multiplying the Fourier blocks, and therefore
the classical limit of this family of operators is the same as that of the simple shift operator: a coherent 
sum of two baker maps. However, the reason the fidelity is unity for all time is due to a rather intriguing if
simply verifiable identity. We will write $F_{2L}$ for $F_{2L}(0,0)$ below. The perturbation operator is:
\beq
V(\theta)=F_{2L}^{-1}\left(\exp(-i \theta \sigma_x) \otimes I_L\right)F_{2L}=I_L\otimes \exp(-i \theta \sigma_z).
\eeq
The last equality is an identity valid for all integer $L$. This in turn simply  follows from the identity: 
\beq
\label{identX}
F_{2L}^{-1} \left( \sigma_x \otimes I_L\right)F_{2L}=  \left( I_L \otimes \sigma_z\right)
\eeq
which maybe directly verified. Note that since $\sigma_x\otimes I_L$ is a circulant matrix it has
to be diagonalized by the Fourier transform, and since the eigenvalues are $\pm 1$ these are
the only possible diagonal entries. It is also easily verified that this has the structure of  $L$ repetitions
of $(1,-1)$ pairs which are the diagonal entries of $\sigma_z$. In some sense the Fourier transform is
simultaneously performing a bit reversal and a ninety degree rotation in qubit space. However note that 
for this identity to be true we do not require that $L$ be a power of $2$. We state here that similar 
identities do {\it not} hold for the other two Pauli matrices, but there are approximations that we will state 
further ahead. Since $\sigma_z$ merely changes the phase of the state (in standard basis) it follows
that the fidelity $f(t)=1$ always.

\subsection{Case: $P=\sigma_y$}

In this case we see that the classical limit is altered by the perturbation. The Hadamard transform
in the qubit space is further rotated around the $y$-axis in spin space and the final operator
is similar to that used in Eq.~(\ref{alhalfX}), which we recall is for the case when the phase
$\al=1/2$ and for a $\sigma_x$ perturbation.
\beq
\label{alzeroY}
S(\theta; 0,\sigma_y)=F_{2L}^{-1} \left( \begin{array}{cc} \sin(\f{\pi}{4}-\theta) F_{L}(0,0)& 
\cos(\f{\pi}{4}-\theta) F_{L}(0,\f{1}{2}) \\  \cos(\f{\pi}{4}-\theta) F_{L}(0,0)&
-\sin(\f{\pi}{4}-\theta) F_{L}(0,\f{1}{2})
\end{array} \right).
\eeq
Therefore this case is the closest to the parity preserving case we have already discussed 
and shown the sharp transitions to features of a quantum chaotic spectrum. In Fig.~({\ref{fid00y})
we plot the fidelity $f(t)$ for a set of $N$ values that are close to 256. We notice immediately that 
although the $N$ values are as close as can be ($N$ must be even) the fidelity decays in qualitatively
different manners. Except for a very short-time scale the decays are different and one sees a prominent 
``shoulder'' in each of the curves at which the fidelity starts to decay even faster. 

The easiest case to discern this in the figure is for $N=254$ when the shoulder occurs at $t=110$.
It is quite easy to numerically relate the time at which this occurs to the multiplicative order of $2$ modulo $N-1$, 
referred to henceforth loosely as simply the order of $N-1$. Recall that this is the smallest number $k_0(N-1)$ such 
that $2^{k_0(N-1)}=1 \;\mbox{modulo} (N-1)$. We are guaranteed that such a number exists because $N-1$ is
an odd integer. Indeed $k_0(253)=110$, while $k_0(255)=8$, $k_0(251)=50$, and $k_0(249)=82$. Thus the fidelity
shows a shoulder either exactly at $t=k_0(N-1)$ or at $t=k_0(N-1)/2$, the first case is observed for
$N=256$ and $N=254$ while the latter is the case for $N=252$ and $250$. It is significant then that for
numbers of larger orders the fidelity can decay considerably even for small perturbations. Note that $ \theta=0.05$
in the figure which roughly translates to a $0.51:0.49$ mixture of the two types of bakers, while a $0.5:0.5$ ``mixture''
will be the unperturbed shift operator. The larger the order is the higher powers of $S$ must be calculated and 
the higher chance of the fidelity to be lowered. It is interesting that the objective of the Shor algorithm
namely finding the order already appears in the fidelity as a critical time.
\begin{figure}
\includegraphics[height=5in,angle=-90]{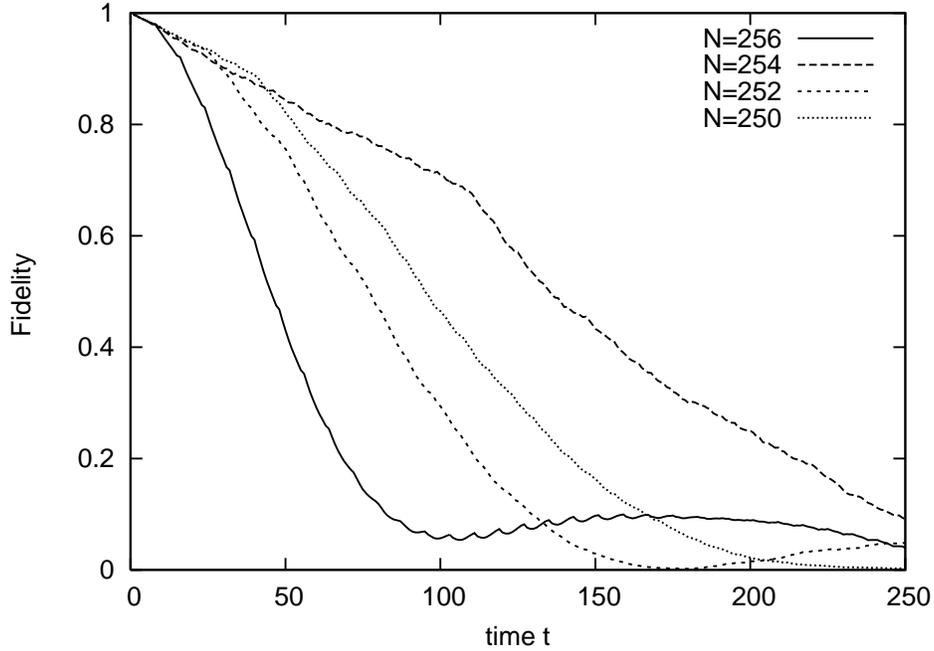}
\caption{The fidelity decay for four neighboring values of $N$. The perturbation is $P=\sigma_y$, the
phase $\alpha=0$, and $\theta=0.05$.}
\label{fid00y}
\end{figure}

We can gain a qualitative understanding of these behaviors with a surprisingly simple model. Consider the
case when $N$ is a power of $2$ say $N=2^M$ and let the perturbation be 
\beq
V(\theta)=I_L\otimes \exp(-i \theta \sigma_x).
\eeq
Note that this has the same structure as the perturbation from the previous case, but is not the 
true perturbation in this one. Then the initial state is $|0\cdots01\kt$ and it is clear that 
for $t\le M$:
\beq
\left(V(\theta) S\right)^t|0\cdots01\kt=I_{2^{M-t}}\otimes \exp(-i \theta \sigma_x) \otimes \exp(-i \theta \sigma_x) \cdots
\otimes \exp(-i \theta \sigma_x) |0\cdots01\kt.
\eeq
and $f(t)=|\cos^2(\theta)|^{t}$. Thus the initial fidelity decay is in fact exponential with a rate
$-\log(|\cos^2(\theta)|)$. However beyond $t=M$ there is an additional error adding up and so 
$f(t)=|\cos^2(\theta)|^{2M-t} |\cos^2(2\theta)|^{t-M}$ for $M<t\le 2M$. Thus in this range of time
the fidely decays about four times as fast. We can write in general for this model that 
\beq
\label{analform}
f(t)=|\cos^2(r \theta)|^{(r+1)M-t} |\cos^2((r+1)\theta)|^{t-rM}
\eeq
where $r=[t/M]$ and $[x]$ is the integer part of $x$. 
We have shown in Fig.~(\ref{fid00y256}) how good an approximation this can be for the case 2 situation when 
$N$ is indeed a power of 2 (or the number we want to factor is one less than a power of 2). There is even good
quantitative agreement. 
\begin{figure}
\includegraphics[height=4in,angle=-90]{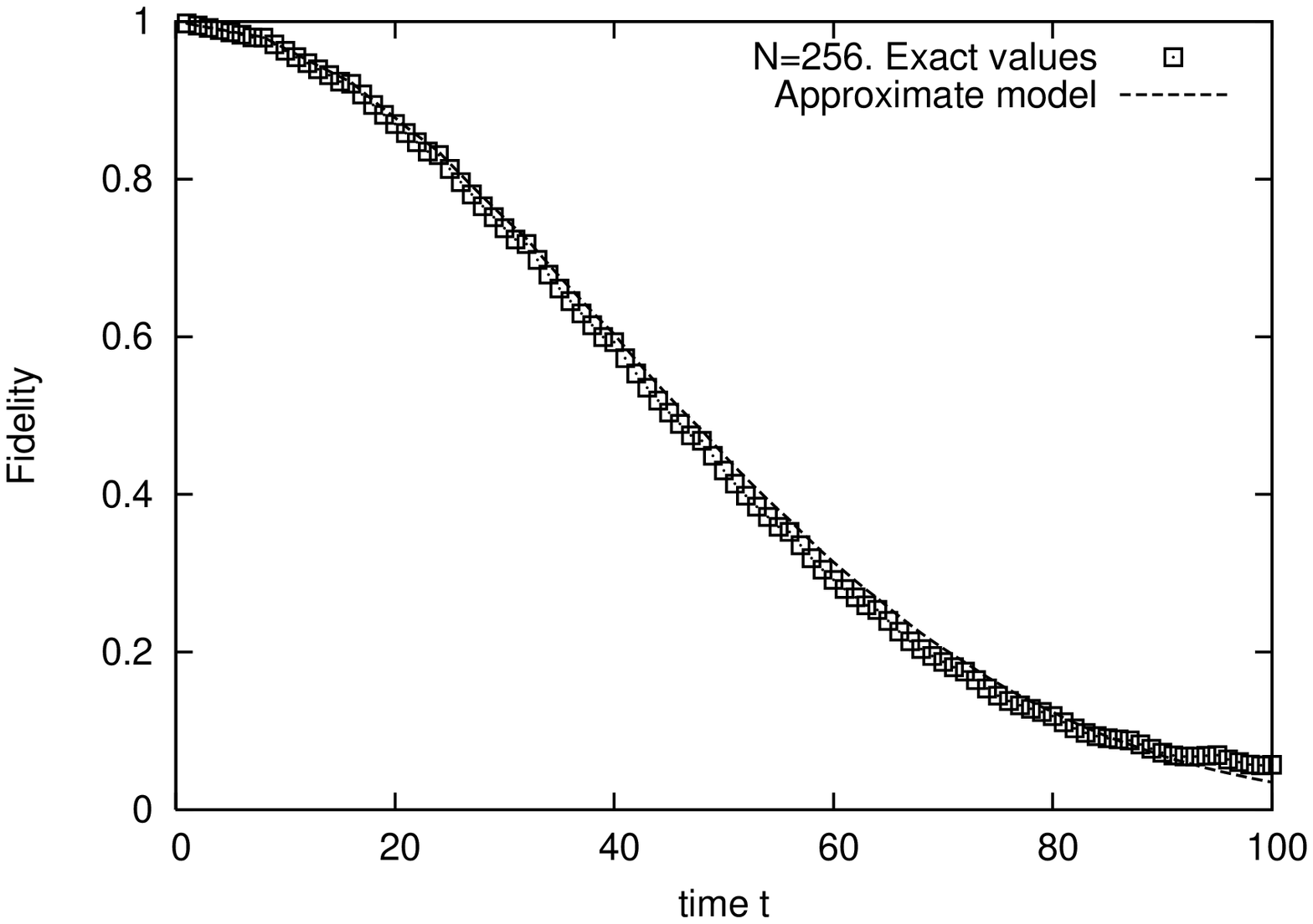}
\includegraphics[height=4in,angle=-90]{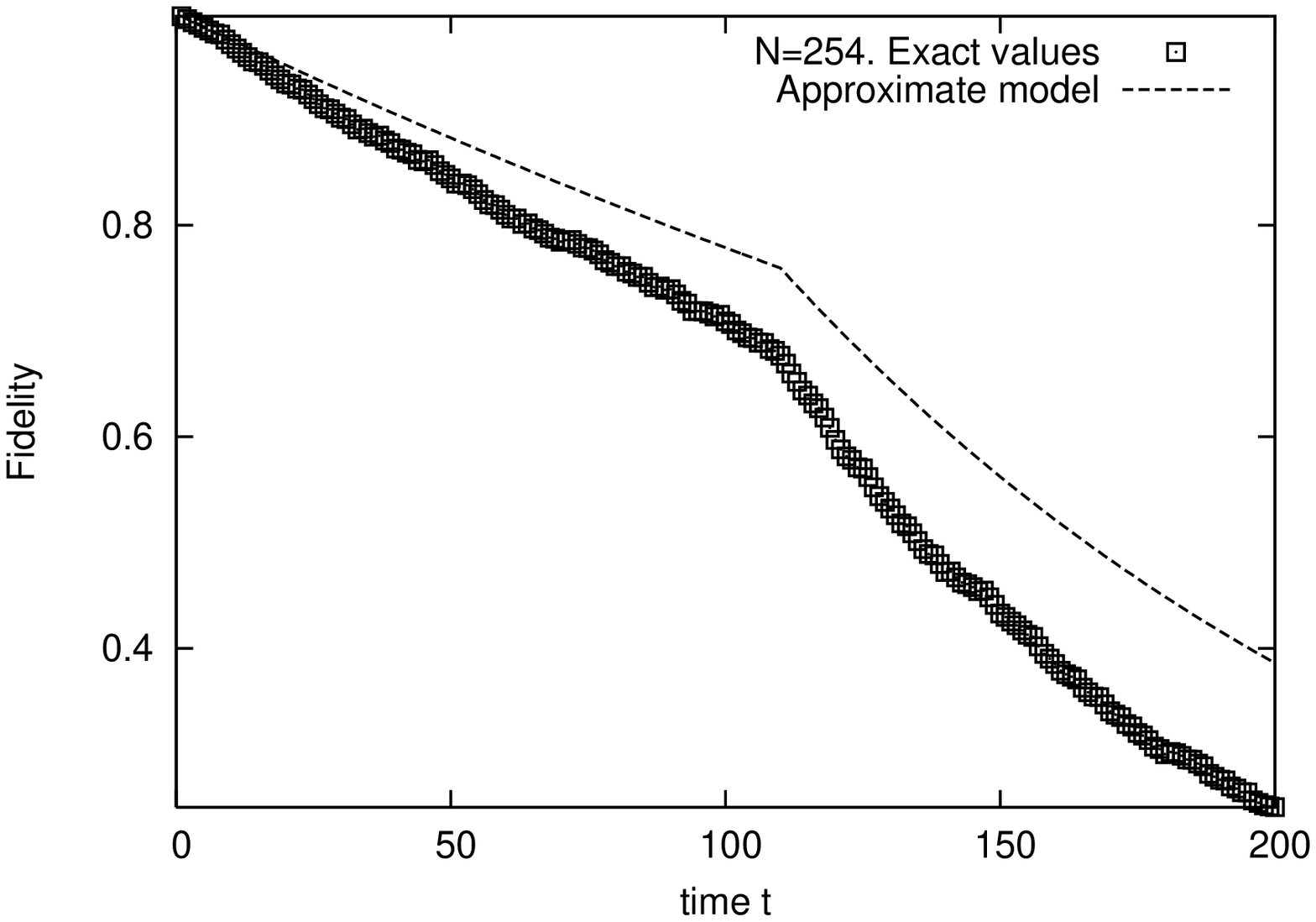}
\caption{The comparison of the fidelity decay for the $\sigma_y$ perturbation ($\alpha=0$, and $\theta=0.05$)
with the analytical estimate in Eq.~(\ref{analform}) from an approximate model for two cases of $N$} 
\label{fid00y256}
\end{figure}
On the other hand when $N=254$, ($253$ has a high order of $110$) the approximate formula is only qualitatively correct
as seen in Fig.~(\ref{fid00y256}).

The model works reasonably well because 
\beq
\label{identexpY}
F_{2L}^{-1} \left( \exp(-i \theta \sigma_y ) \otimes I_L\right)F_{2L} \approx  \left( I_L \otimes \exp(-i \theta \sigma_x)\right),
\eeq
which follows from
\beq
\label{identY}
F_{2L}^{-1} \left( \sigma_y \otimes I_L\right)F_{2L} \approx  \left( I_L \otimes \sigma_x\right),
\eeq
so that exact perturbation operator which would have been the L.H.S. of Eq.~(\ref{identexpY}) is approximated by its
 R.H.S. which we have used above. This is the counterpart of Eq.~(\ref{identX}), however here this is only an approximation. 

This ``model'' or approximation does not explain the appearance of half of the periods for
some values of $N$, such as for $252$ and $250$ above. Indeed when $N$ is a power of $2$ we
will always observe the first shoulder at the order of $N-1$. This is in fact the result of the possibility that 
there exists an integer $k'$ such that $2^{k'}\, \mbox{mod}\, (N-1)=-1$, which implies that $k'=k_0/2$.
If there exists such an integer then we must, according to Shor's algorithm, choose a different integer (other than 2) to find its order of. That is we cannot use the order of $2$ to find a factor of $N-1$, which is the ultimate objective. For our analysis, this situation implies that 
\beq
\label{spowhalf}
S^{k_0(N-1)/2}\equiv R'=\left(\begin{array}{ccc}1&&\\&R_{N-2}&\\&&1\end{array}\right)
\eeq
where $R_{N-2}$ is the parity operator with $1$ along its secondary diagonal and zero elsewhere.
That is $S^{k_0(N-1)/2}$ is almost the parity operator except that $0$ and $N-1$ instead of being 
interchanged are fixed by $S$, and hence all its powers. To clarify Eq.~(\ref{spowhalf}) may or
may not hold depending on if $2^{k_0(N-1)/2}=-1\, \mbox{mod}\, (N-1)$ or not. For instance this
is never the case if $N$ is a power of 2. We note that this simplified model has also been considered
by \cite{Scott07} recently to show that exponential fidelity decay does not necessarily mean an
hypersensitivity to perturbations. However in the context of this paper it is interesting that the
model works approximately even when $N$ is not a power of $2$ and preliminary results indicate
 that there is hypersensitivity to perturbations as well \cite{Unpublished}.

In general for an initial state $|\psi_0\kt$ we have that 
\beq
f(t)=|\br \psi_0|V_t V_{t-1}\cdots V_1|\psi_0\kt|^2
\eeq
where $V_l$ is the perturbation in the interaction picture: $V_l=S^{-l}VS^{l}$. Thus always $V_{k_0(N-1)}=V$
but in the situation where Eq.~(\ref{spowhalf}) holds we have that $V_{k_0(N-1)/2}=R'^{-1}VR'$. Thus it is 
clear that these times are special for the fidelity as seen in the numerical calculations as well. While these arguments
along with the approximate model gives a fair understanding of the decay, it is not
complete  and a more detailed analysis of the product above must be carried out, which
the author is unable to provide. 
There is a rather large literature surrounding the so-called Loschmidt echo \cite{FidelReview}, or fidelity, wherein quantum chaotic systems have been subjected to a small perturbation on reversal. The current discussion is in fact closely related,
 but previous work has naturally concentrated on the generic case of a non-degenerate operator that is perturbed. In the case
of the shift operator, it can be highly degenerate, as well as completely periodic, thereby making
it ``non-generic''. It can be compared again to the quantum chaotic cat maps that are also periodic and
degenerate in general. Smooth perturbations of this for instance of the type that has been studied 
before \cite{PertCat} could produce fidelity decays of a similar character that we have noted here.

\subsection{Case: $P=\sigma_z$}
In this case 
\beq
S(\theta; 0,\sigma_z)=F_{2L}^{-1}(0,0) \left( \begin{array}{cc} \f{1}{\rt}e^{-i \theta} F_{L}(0,0)& 
\f{1}{\rt}e^{-i \theta} F_{L}(0,\f{1}{2}) \\ \f{1}{\rt}e^{i \theta} F_{L}(0,0)&
 \f{1}{\rt}e^{i \theta}  F_{L}(0,\f{1}{2})
\end{array} \right).
\eeq
Note that there seems to be only a minor change, namely those of signs of phases in the Fourier blocks,
compared to the first case, and also the classical limit still  remains unaltered by the phase-flip 
perturbation. However,  the fidelity does decay even due to the ``quantum perturbation'' and seems 
to be of a similar character to that observed when $P=\sigma_y$, namely the previous case. The differences
start showing up sharply in the case when Eq.~(\ref{spowhalf}) holds, namely when $k_0(N-1)$ is such that 
$2^{k_0(N-1)}=-1\, \mbox{mod}\, (N-1)$. It appears to be generically the case that beyond this time there
are large oscillations reminiscent of fidelity decay in near-integrable systems \cite{Sankar}. This is illustrated in Fig.~(\ref{fid00z}) for values of $N$. In the cases when $N=252$ and $250$ Eq.~(\ref{spowhalf}) holds and we see that 
beyond time of half the order there are regular oscillations with this period.

That we must expect a fidelity decay is due to a counterpart of the approximation used in the
previous case, namely
\beq
\label{identexpZ}
F_{2L}^{-1} \left( \exp(-i \theta \sigma_z ) \otimes I_L\right)F_{2L} \approx  \left( I_L \otimes \exp(-i \theta \sigma_y)\right),
\eeq
which follows from
\beq
\label{identZ}
F_{2L}^{-1} \left( \sigma_z \otimes I_L\right)F_{2L} \approx  \left( I_L \otimes \sigma_y\right),
\eeq
which is a result of combining the identity in Eq.~(\ref{identX}) and the approximation in 
Eq.~(\ref{identY}).
\begin{figure}
\includegraphics[height=5in,angle=-90]{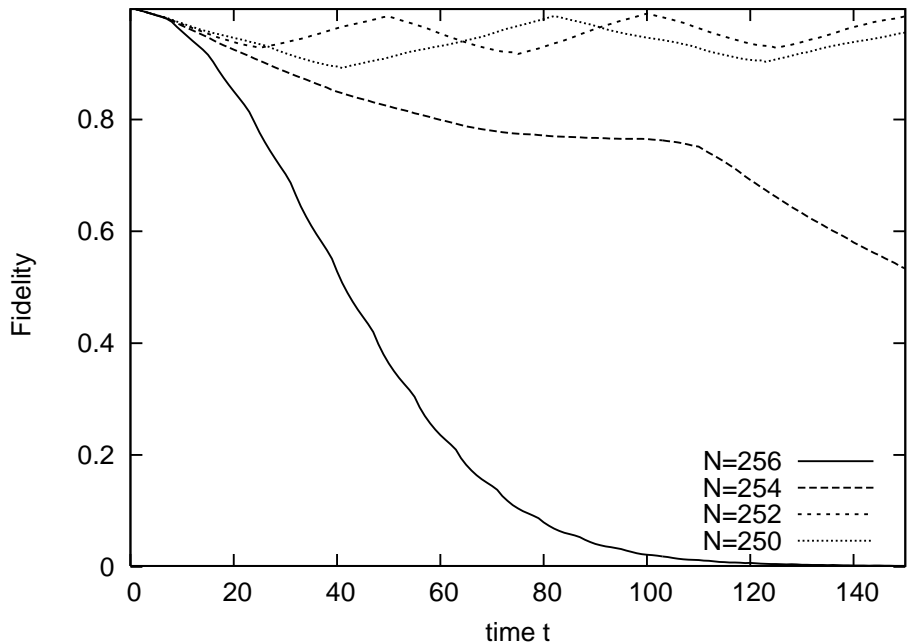}
\caption{The fidelity decay for four neighboring values of $N$. The perturbation is $P=\sigma_z$, the
phase $\alpha=0$, and $\theta=0.05$.}
\label{fid00z}
\end{figure}
Thus this final case of perturbation we consider is sort of intermediate between cases A and B, however for
practical purposes it is closer to case B, as the time behavior beyond the time of the order or half the order
is not likely to be of interest from the point of view of the Shor algorithm. The approximate formula in
Eq.~(\ref{analform}) continues to be approximately good for those $N$ for which the condition in Eq.~(\ref{spowhalf})
does not hold, and for those for which it does, it is approximately good till time of half the order at which
the oscillations begin. 

\section{Discussion}

We have studied three archetypal perturbations, phase-flip, bit-flip and a combination therefore,
 that are possible in the critical part of Shor's
algorithm, namely the modular multiplication or exponentiation part. We have confined ourselves to the simplest
possible case when the multiplier is 2, when these perturbations can be interpreted in terms
of coherent superpositions of quantum bakers maps, whose classical limits are completely chaotic,
and are models of randomness. Thus we have shown that there are generic perturbations of the
modular exponentiation operator that will qualify as ``quantum chaotic''. We have shown this by 
computing the nearest-neighbor  spacing statistics and seeing that it is of the type expected of
random matrices. More pertinent to the algorithm itself we have studied the fidelity decay 
that occurs with the relevant initial state and shown that for the three types of perturbations there
are three possible fidelity decay behaviors. This can be interpreted in terms of the fact that 
some perturbations alter the classical limit while some do not, as well as in arising
from some identities (one exact and one approximate) that involve the Pauli spin matrices and the Fourier transform, which
while the author has not seen before, are completely elementary and likely to be known and useful already.
A simple model of the fidelity decay is afforded by these identities that describes surprisingly well the
exponential decay in time punctuated by shoulders at times related to the order. An exact solution of the problem seems unlikely, and semiclassical analysis cumbersome due to the fact that  the modular exponentiation (when the multiplier is 2) is essentially the {\it sum} of two unitary operators with well defined classical limits. 

The precise impact of the exponential fidelity decay on the functioning of the algorithm remains to be seen. Such an 
study for the case of static imperfections was recently carried out \cite{Shepel07}. We have been 
primarily interested in pointing to the deep and exact relationship between the modular exponentiation
part of the Shor algorithm and the quantization of an archetypal model of classical deterministic
chaos, namely the bakers map. If we had larger multipliers that 2, as will generally be the case, it is 
reasonable to expect that these will be related to generalized bakers maps with more than 2 partitions \cite{BalVor2}. The
number of possible stacking are also more, but it is completely conceivable that once again there 
are perturbations to the modular exponentiation operator that are quantum chaotic and close to the
quantization of such bakers. That such bakers will have larger Lyapunov exponents and have greater
classical randomness may make the quantum operators even more susceptible to such gate errors;
however this is at the moment mere speculation. 

\acknowledgments{It is a pleasure to thank Steven Tomsovic and Alfredo M Ozorio de Almeida for 
discussions}


\begin{thebibliography}{99}
\bibitem{Voros} This terminology is due to the quantum baker Andr\'e Voros.
\end{thebibliography}
\end{document}